\documentclass[aps,prb,twocolumn,groupedaddress,showpacs,floatfix,altaffilletter]{revtex4-1}
\usepackage{graphicx}
\usepackage{grffile}
\usepackage{amsmath}
\usepackage{amssymb}
\usepackage{bm}
\usepackage{color}
\usepackage[dvipsnames]{xcolor}
\usepackage{amsmath,amssymb,amsfonts}
\usepackage{epsfig}
\usepackage{times}
\usepackage{inputenc}
\usepackage[colorlinks,bookmarks=false,citecolor=blue,linkcolor=red,urlcolor=blue]{hyperref}

\newcommand{\eref}[1]{Eq.~(\ref{#1})}

\begin{document}

\title{Potential energy contribution to the thermopower of correlated electrons}

\author{R.~Nourafkan}
\email{reza.nourafkan@usherbrooke.ca} 
\author{A.-M.S. Tremblay}
\affiliation{$^1$D{\'e}partement de Physique, Institut quantique and RQMP, Universit{\'e} de Sherbrooke, Sherbrooke, Qu{\'e}bec, Canada J1K 2R1}
%
%
\begin{abstract}
Certain classes of strongly correlated systems promise high thermopower efficiency, but a full understanding of correlation effects on the Seebeck coefficient is lacking. This is partly due to limitations of Boltzmann-type approaches. One needs a formula for the thermopower that allows separate investigations of the kinetic and potential energy contributions to the evolution with temperature and doping of the thermopower. Here we address this issue by deriving for Hubbard-like interactions a formula for the thermopower that separates the potential from the kinetic energy contribution and facilitates a better understanding of correlation effects on the Seebeck coefficient. As an example, the thermopower of the one-band Hubbard model is calculated from dynamical mean-field. For  interactions in both the intermediate and strong correlation limit, the contributions from kinetic and potential energy nearly cancel.   
\end{abstract}

\maketitle
\section{Introduction}
The Seebeck coefficient, or thermopower, measures the magnitude of an induced voltage in response to a temperature difference. Materials with large thermoelectric power 
are of great economic and environmental interest and can be utilized in a wide range of applications, most notably for waste-heat recovery. Hence, an efficient power generation technology based on thermoelectric materials could be an important part of the solution to today's energy challenge.  The key, though, is finding thermoelectric materials with much higher conversion efficiency than currently known thermoelectric  materials. 
To this end, strongly correlated electron systems have shown promises. Indeed, 
it has been shown experimentally that some materials, such as sodium cobalt oxide Na$_x$CoO$_2$,~\cite{PhysRevB.56.R12685} or narrow-gap semiconductors, such as FeSb$_2$ (see~\cite{Tomczak_2018} and references therein) possess unusually large thermopower, due in part to strong electron interactions. 

Fundamentally, thermoelectric current arises from particle-hole (p-h) asymmetry. Electron interactions can influence p-h asymmetry in several ways: By altering the electronic structure, by modifying the current matrix elements, or by introducing  frequency- and momentum-dependent scattering rates (relaxation times).  This provides multiple reasons to investigate systems with strong correlations to search for large thermopower. 
One way to expand our understanding of interaction effects on thermopower is to investigate how potential energy contributes to thermopower. Such a contribution may be measured experimentally in a controlled setting, such as cold atom systems that spatially-resolve double occupancy.~\cite{Brantut713, Esslinger_2008,bakr2010probing, Brown379}

At low-temperatures, electrical/heat currents are carried by low-energy excitations. In weakly correlated systems, these excitation are long-lived quasiparticles, allowing to use simple versions of the Kubo formalism or semi-classical Boltzmann theory~\cite{Mahan_Sofo_1996} to simulate transport properties. 
These approaches have been successful for conventional  thermoelectric  materials, which happen to be weakly correlated semiconductors.  In  these systems, the heat carried by excitation is $\epsilon_{\bf k}-\mu$, where $\epsilon_{\bf k}$ denotes single electron eigenenergy and $\mu$ is the chemical potential.  Therefore, this approach only accounts for the kinetic energy contribution to thermopower.  Boltzmann transport theory cannot be used for strongly correlated materials because the underling assumptions, such as well-defined quasi-particles, do not hold and also because it neglects the potential energy contribution to thermopower.

The formula that is currently used for thermopower of correlated electrons is based on an extension of the Kubo formalism for non-interacting systems: the heat carried by particle-hole excitations is $\omega$, their excitation energy measured with respect to chemical potential. Interaction effects appear clearly only in the width of spectral functions.~\cite{PhysRevB.67.115131, PhysRevLett.111.036401, PhysRevLett.109.017001, Deng_Mravlje_zitko_Ferrero_Kotliar_Georges_2013} Although, this formula is usually derived for non-interacting systems and then extended to interacting systems, one can argue that it should be an exact formula for interacting systems as well since $\omega$ is the many-body excitation energy of the p-h. Here, we show that this  is indeed the case by deriving many-body energy/heat current operators and a formula for the electronic part of the Seebeck effect that accounts for both kinetic and potential energy contributions to thermopower. Our formula allows to investigate separately interaction effects on thermopower in a clearer and more quantitative manner.

\section{Potential and Kinetic Energy Contribution in Thermopower}
Transport coefficients like the electrical conductivity $\sigma$, and the thermoelectric power $S$ are related to response functions $L^{ij}_{\alpha\beta}$ defined as
\begin{equation}
L^{ij}(i\nu_n)=\frac{1}{\beta V}\int_0^{\beta}e^{i\nu_n\tau}\langle T_{\tau} {\bf J}_i(\tau)\cdot {\bf J}_j(0) \rangle.
\end{equation}
There ${\bf J}_i^{\alpha}$ is the $\alpha$ component of the particle current ($i=1$) or thermal current ($i=2$), $V$ is the volume, $\beta=1/k_BT$, and $T_{\tau}$ denotes the time ordering operator. The retarded response functions that determine the transport coefficients are obtained by analytic continuation $i\nu_n \rightarrow \nu+i0^+$ from $\mathcal{A}^{ij}(\nu)\equiv \Im L^{ij}(\nu)/\nu$.  In particular, the frequency-dependent electrical conductivity and the Seebeck coefficient are given, respectively,  by $\sigma=e^2\mathcal{A}^{11}(\nu)$ and $S(\nu)=-(k_B/eT)\mathcal{A}^{12}(\nu)/\mathcal{A}^{11}(\nu)$, where $k_B$ is Boltzmann's constant, and $e$ is the elementary charge.  The DC limit is obtained by taking the limit $\nu=0$.



Using exact forms for current and heat operators of  a one-band model with Hubbard interaction,  we obtain the following equation for $L^{12}(i\nu_n)$ \footnote{The appendix contains a detailed derivation of the thermopwer formula, along with its analytic continuation.}
\begin{widetext}
\begin{align}
L^{12}_{\alpha\alpha}(i\nu_n)\simeq -\frac{1}{2\beta V}\sum_{{\bf k}\sigma}\sum_{\omega_m}
&\bigg( {\bf v}^{\alpha}_{{\bf k}}G_{{\bf k}\sigma}(i\omega_m+i\nu_n)\Lambda_{\bf k}(i\omega_m, i\nu_n){\bf v}^{\alpha}_{{\bf k}} G_{{\bf k}\sigma}(i\omega_m) + (\nu_n \leftrightarrow -\nu_n)
\bigg)\nonumber\\
\Lambda_{\bf k}(i\omega_m, i\nu_n)&\equiv \epsilon_{\bf k}-\mu+\Sigma_{{\bf k}\sigma}(i\omega_m+i\nu_n)/2+\Sigma_{{\bf k}\sigma}(i\omega_m)/2 ,\label{L12}
\end{align}
\end{widetext}
where $\Lambda_{\bf k}(i\omega_m, i\nu_n){\bf v}^{\alpha}_{{\bf k}}$ is the heat-current vertex; as seen from the explicit form of $\Lambda_{\bf k}(i\omega_m, i\nu_n)$, the interaction effects on the energy vertex is captured by adding a suitable  portion of the self-energy to the kinetic energy. This is similar to what we have shown for  orbital magnetization.~\cite{PhysRevB.90.125132}  
In ~\eref{L12}, $G_{{\bf k}\sigma}(i\omega_m)$ denotes the interacting propagator and  ${\bf v}^{\alpha}_{{\bf k}}=\partial \epsilon_{\bf k}/\partial k_{\alpha}$ the velocity. The vertex correction effects can be included by replacing ${\bf v}^{\alpha}_{{\bf k}}$ in the heat-current vertex by $-\partial G^{-1}_{{\bf k}\sigma}/ \partial k_{\alpha}$. We do not address the question of vertex corrections in this paper.
The interacting propagator is given by
\begin{equation}
G_{{\bf k}\sigma}(i\omega_m)=[i\omega_m+\mu-\epsilon_{\bf k}-\Sigma_{{\bf k}\sigma}(i\omega_m)]^{-1},
\end{equation}
where $\epsilon_{\bf k}$ denotes non-interacting electron dispersion and $\Sigma$ is the electron self-energy.  

Equation~\ref{L12} is equivalent to the currently used formula for thermopower,~\cite{PhysRevB.67.115131, PhysRevLett.111.036401, PhysRevLett.109.017001} at least in the limit where vertex corrections can be neglected. This can be seen as follows: the heat vertex in \eref{L12} can be rewritten as 
\begin{equation}
    [i(\omega_m+\nu_n)-G^{-1}_{{\bf k}\sigma}(i\omega_m+i\nu_n)]/2+[i\omega_m-G^{-1}_{{\bf k}\sigma}(i\omega_m)]/2.
\end{equation}  
Upon replacing this form of the heat vertex in the equation for $L^{12}$, terms coming from inverse Green's functions simplify one of the two Green's function so that the remaining terms proportional to $G_{{\bf k}\sigma}(i\omega_m)$  can be neglected after Matsubara frequency summation and analytic continuation. Indeed, they are purely real and frequency independent. Therefore, the above equation is equivalent to the currently used formula that uses $i(\omega_m+\nu_n/2)$ for  the heat vertex.~\cite{PhysRevB.67.115131}   However, our formula allows us to separate kinetic and potential energy contributions to the Seebeck effect, and to study how each contribution is impacted by temperature and doping. 
Furthermore, although we do not take into account  vertex corrections here, we would like to emphasize that the two formalisms may yield different results once vertex corrections are taken into account. In this case, one cannot simply neglect the contributions from the inverse of the Green's function. However, considering vertex corrections is beyond the scope of this paper~\footnote{In particular, high-frequency sum rules~\cite{PhysRevB.73.085117} will not be satisfied without vertex corrections.}. 

After analytic continuation, we obtain the following expression for $\mathcal{A}^{12}(\nu)$ \cite{Note1}
\begin{widetext}
\begin{align}
\mathcal{A}^{12}(\nu)=&-\frac{\pi}{ V}\sum_{{\bf k}\sigma}\int d\omega
\bigg( {\bf v}^{\alpha}_{{\bf k}}A_{{\bf k}\sigma}(\omega+\nu)[\epsilon_{\bf k}-\mu+\Re \Sigma_{{\bf k}\sigma}(\omega+\nu)/2+\Re \Sigma_{{\bf k}\sigma}(\omega)/2]{\bf v}^{\alpha}_{{\bf k}} A_{{\bf k}\sigma}(\omega)
\bigg)\frac{(n_F(\omega+\nu)-n_F(\omega)}{\nu})
\nonumber\\
&-\frac{\pi}{2V}\sum_{{\bf k}\sigma}\int d\omega \bigg( {\bf v}^{\alpha}_{{\bf k}}\Re G_{{\bf k}\sigma}(\omega+\nu)B_{{\bf k}\sigma}(\omega+\nu) {\bf v}^{\alpha}_{{\bf k}} A_{{\bf k}\sigma}(\omega)\bigg)\frac{(n_F(\omega+\nu)-n_F(\omega)}{\nu})\nonumber\\
&-\frac{\pi}{2V}\sum_{{\bf k}\sigma}\int d\omega \bigg( {\bf v}^{\alpha}_{{\bf k}}A_{{\bf k}\sigma}(\omega+\nu)B_{{\bf k}\sigma}(\omega) {\bf v}^{\alpha}_{{\bf k}} \Re G_{{\bf k}\sigma}(\omega)\bigg)\frac{(n_F(\omega+\nu)-n_F(\omega)}{\nu}),\label{A12}
\end{align}
\end{widetext}
where we defined the spectral weights for the Green's function and self-energy respectively by $A_{{\bf k}\sigma}(\omega)=(-1/\pi)\Im G_{{\bf k}\sigma}(\omega)$ and $B_{{\bf k}\sigma}(\omega)=(-1/\pi)\Im \Sigma_{{\bf k}\sigma}(\omega)$. The quantity $n_F$ is the Fermi function.~\footnote{ Our numerical results show that equation~\ref{A12} and the conventional formula for $\mathcal{A}_{12}$ give identical results for vanishingly small broadening.} $\mathcal{A}^{11}(\nu)$ is given by a formula similar to the first line of Eq.~\ref{A12}, except that the expression in square bracket is replaced with the identity.~\cite{PhysRevLett.109.017001}

It is natural to define the kinetic and the potential energy contributions to the Seebeck effect by decomposing the energy vertex as, $\Lambda_{\bf k}\equiv \Lambda^{K.E.}_{\bf k}+\Lambda^{P.E.}_{\bf k}$, where, measuring energy with respect to the renormalized chemical potential, we find 
\begin{align}
\Lambda^{K.E.}_{\bf k}&\equiv \epsilon_{\bf k}-\mu+\Re\Sigma_{{\bf k}\sigma}(0),\label{decompKin}\\
\Lambda^{P.E.}_{\bf k}&\equiv \Sigma_{{\bf k}\sigma}(i\omega_m+i\nu_n)/2+\Sigma_{{\bf k}\sigma}(i\omega_m)/2 -\Re\Sigma_{{\bf k}\sigma}(0).\label{decomp}
\end{align}
The corresponding terms in the analytically continued expression are rather easy to identify by focusing first on the kinetic energy part \eref{decompKin}. 
Although it is not possible to measure these contributions individually, except perhaps in cold-atom experiments that can access spatially-resolved double occupancy ~\cite{Esslinger_2008, bakr2010probing}, their theoretical study allows a deeper understanding of their interplay, as we shall see. We move on to applications of this formula.

\section{Results}
\subsection{Model}
We apply Eq.~\ref{L12} and Eq.~\ref{A12} to a system described by the single-band Hubbard model as an example that will illustrate the separate effects of kinetic and potential energy on thermopower. 
The Hamiltonian  on a square lattice reads
\begin{equation}
H=-\sum_{ij,\sigma}t_{ij}c^{\dagger}_{i\sigma}c_{j\sigma}+U\sum_i n_{i\uparrow}n_{i\downarrow}.
\end{equation}
We take the values of the hopping parameters for lanthanum copper oxyde from Ref.~\cite{PhysRevLett.121.077004}. The first, second and third nearest-neighbour hopping are $t_1=190$~meV, $t_2=-0.12t_1$ and $t_3=0.06t_1$. We consider two values for Hubbard interaction: $U=6t_1$ and $U=16t_1$. For $U=6t_1$, system is moderately correlated while for $U=16t_1$ it undergoes a phase transition to a Mott phase at half-filling. We solve this model using dynamical mean field theory (DMFT) ~\cite{RevModPhys.68.13} with an exact-diagonalization solver~\cite{PhysRevLett.72.1545, PhysRevB.80.155130, PhysRevB.82.115127}.  
We investigate the Seebeck effect as a function of temperature and hole-doping for these two cases.

\subsection{General Considerations} In normal metals, the Seebeck coefficient scales as $S ~\simeq  (k_B T/E_F)(k_B/e)$, where $E_F$ denotes the Fermi energy. Hence, the thermopower of normal metals is linear in temperature and vanishes at $T=0$. In semiconductors, however, both entropy and the density of the mobile carriers (consequently the electrical conductivity) vanish at zero temperature and the Seebeck coefficient is determined by the relative rate of decrease in these two vanishing quantities and can acquire large values upon approaching $T=0$.~\cite{PhysRevLett.116.087003}  Similarly, the electrical conductivity of a lightly doped Mott system depends non-trivially on temperature. Hence, one way to obtain a large Seebeck coefficient is to design a strongly correlated system in which electrical conductivity decreases faster than entropy upon decreasing temperature. This might be  achievable for a lightly-doped Mott system, at least, at a finite range of doping values. 

As mentioned earlier, the Seebeck coefficient is sensitive to the particle-hole asymmetry around the chemical potential in the spectral functions (or density of states), in the current matrix elements (velocity matrices) and in the energy-dependent scattering rates.~\cite{ doi:10.1143/JPSJ.76.083707, PhysRevMaterials.3.075404, PhysRevB.78.115121, Deng_Mravlje_zitko_Ferrero_Kotliar_Georges_2013, PhysRevB.87.035126} These quantities are not independent of each other. Consider our model in the absence of interaction. The density of states has a van-Hove singularity at electron density $n\simeq 0.8$ or at hole doping $p\simeq 0.2$.~\cite{PhysRevLett.121.077004} For electron densities smaller than $0.8$,  the chemical potential, $\mu$, lies below the van-Hove singularity, hence, the number of states above the Femi level is much larger than the one below. This inequality reverses for  electron densities larger than $0.8$. Therefore, with constant current matrix elements (velocities) and relaxation time, one expects a sign change in Seebeck coefficient once the chemical potential passes the van-Hove singularity energy. However, velocities are not constant and can partially compensate for the asymmetry in the density of states, shifting Seebeck's sign change to higher electron densities. In the following, in part for this reason, the Seebeck coefficient is negative on a wider range of dopings than expected from the density of states. 
%
\begin{figure}
\begin{center}
\includegraphics[width=0.85\linewidth]{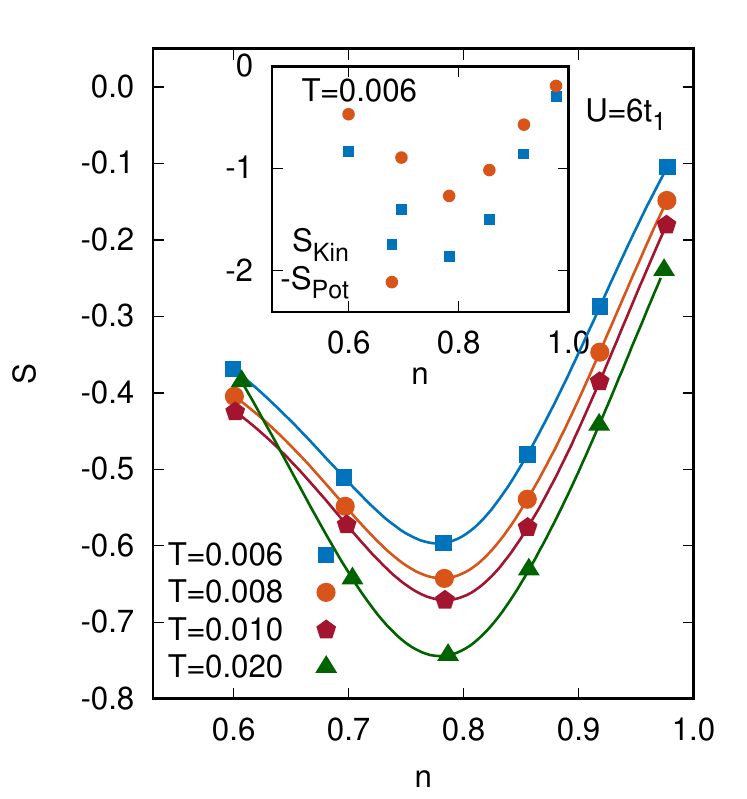} 
\end{center}
\caption{ Seebeck coefficient (in unit of $k_B/e$) as a function of density for several temperatures.  The inset shows the kinetic and the potential contribution to the thermopower at $T=0.006$~eV. The interaction strength is $U=6t_1$. }
\label{fig1}
\end{figure}

\subsection{Intermediate Correlations} Figure~\ref{fig1} shows  the Seebeck coefficient as a function of density for several temperatures and  $U=6t_1$.  It is electron-like with a maximum absolute value around $n=0.8$, resulting from a minimum in electrical conductivity  and a plateau in $\mathcal{A}_{12}$ around this density (not shown).  

The inset of Fig.~\ref{fig1} shows  the kinetic and the potential contributions to the thermopower, as defined in  ~\eref{decompKin} and~\eref{decomp}, as a function of density for $T=6$~meV. Each component is much larger than the Seebeck effect itself, but because they have different signs, their sum is a comparatively small number. 
In this intermediate regime of interactions, $U=6t_1$, and high hole-doping, the kinetic energy contribution is negative and its absolute value is larger than that of the potential energy. Hence, the kinetic energy contribution and the total thermopower have similar signs. Upon approaching half-filling, the absolute value of both components initially increases and then decreases for $n > 0.8$.
The opposite sign of the potential contribution can be understood intuitively: in an electron-like thermopower setup, as we have here, there are more electrons on the cold side, therefore there are more doubly occupied sites there. Due to this density gradient of doubly occupied sites they move toward the warmer side, transferring energy back to that side.

\begin{figure}
\begin{center}
\includegraphics[width=0.85\linewidth]{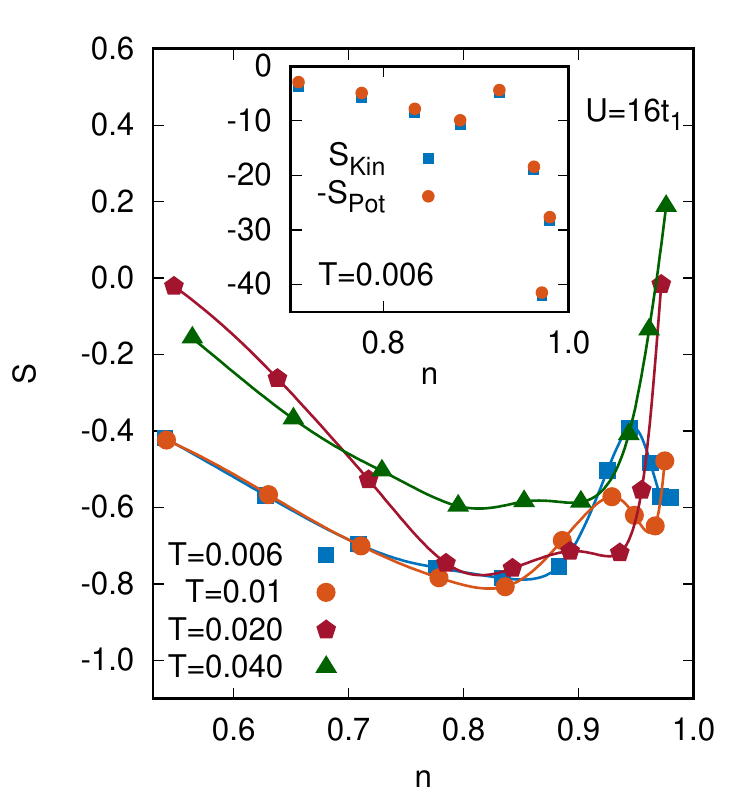} 
\end{center}
\caption{Seebeck coefficient (in units of $k_B/e$) as a function of density for several temperatures.  The inset shows the kinetic and the potential contribution to the thermopower at $T=0.006$~eV. The interaction strength is $U=16t_1$. }
\label{fig2}
\end{figure}

\subsection{Strong Correlations} Figure~\ref{fig2} shows the strongly correlated case $U=16t_1$.  The system is then in a Mott phase at half-filling. At high hole-doping, the thermopower is electron-like and again its magnitude is maximum at densities around $0.8$. Closer to half-filling, the absolute value of the thermopower has a shallow minimum  around doping $p=1-n=\simeq 0.05$ for $T=6$~meV, a minimum that is washed away at larger $T$. At lightly hole-doping region, upon approaching the top of the lower Hubbard band, the thermopower quickly changes sign. Increasing temperature, the sign change happens at larger hole doping (see $T=0.04$ eV, the tendency is present for smaller $T$).  The inset of Fig.~\ref{fig2} shows the kinetic and potential contributions to the Seebeck effect. Similar to the preceeding subsection, at high doping the kinetic energy contribution  determines the sign of the thermopower. The absolute values of these contributions increase upon approaching half-filling before they quickly decrease and change sign, in close vicinity of half-filling (not shown).  The sign changes don't occur at the same density.

\section{Summary}
In summary, we introduced a practical many-body approach for the calculation of the thermopower of interacting systems. Our formula separates the potential energy contribution to the Seebeck effect from its kinetic energy contribution, allowing for physical insight of the interaction effects. We found large cancellations between potential and kinetic energy contributions, with potential energy dominating only close to half-filling in the doped Mott insulator regime. A direct extension of Eq.~\ref{L12} to the transport coefficient $L_{22}$, entering in the definition of the thermal conductivity, can be done. 

\begin{acknowledgments}
R.~N. acknowledge discussion with A. Millis and A. Georges.  This work has been supported by the Natural Sciences and Engineering Research Council of Canada (NSERC) under grant RGPIN-2019-05312, and by the Canada First Research Excellence Fund.  Simulations were performed on computers provided by the Canadian Foundation for Innovation, the Minist\`ere de l'\'Education des Loisirs et du Sport (Qu\'ebec), Calcul Qu\'ebec, and Compute Canada.
\end{acknowledgments}

\appendix
\section{Derivation}
This appendix contains a detailed derivation of the thermopwer formula, along with its analytic continuation.

\subsection{Particle and heat current operators}
In order to calculate the response function $L^{ij}$, we have to find the proper expressions for the particle current $\bf J$ and the energy current ${\bf J}_E$, the latter being related to the heat current ${\bf J}_Q$ by ${\bf J}_Q={\bf J}_E-\mu{\bf J}$ where $\mu$ is the chemical potential.

Consider a real-space representation for the Hamiltonian of a single band Fermi-Hubbard system 
\begin{align}
H=\sum_i h_{i}=\sum_{i{\bm \delta}\sigma}&(t_{{\bm \delta}}/2)(c^{\dagger}_{i+\delta \sigma}c_{i\sigma}+c^{\dagger}_{i \sigma}c_{i+\delta\sigma}) \nonumber\\&+\sum_{i} Un_{i\uparrow}n_{i\downarrow},\label{H}
\end{align}
where ${\bm \delta}$ denotes nearest neighbour bonds. 

\subsubsection{Particle current} The particle current is found as the time derivative  of a particle position operator ${\bf R}_N$ defined as
\begin{align}
{\bf R}_N&=\sum_i {\bf R}_i n_i,\\
{\bf J}&= -i[{\bf R}_N, H]= -i\sum_{ij}  {\bf R}_i [n_i,h_j].
\end{align}
Calculating the commutator using Eq.~\ref{H} yields the following expression for the particle current
\begin{align}
{\bf J}= -i\sum_{i{\bm \delta}\sigma}(t_{{\bm \delta}}/2) ( {\bf R}_{i+{\bm \delta}}-{\bf R}_i)
(c^{\dagger}_{i+{\bm \delta} \sigma} c_{i \sigma}-c^{\dagger}_{i \sigma} c_{i+{\bm \delta} \sigma}). 
\end{align}
Note that $({\bf R}_{i+{\bm \delta}}- {\bf R}_{i})={\bm \delta}$. 

\subsubsection{Energy current} Similarly, the energy current is given by the time derivative of an energy position operator ${\bf R}_E$ defined as
\begin{align}
{\bf R}_E&=\sum_i {\bf R}_i h_i,\\
{\bf J}_E&= -i[{\bf R}_E, H]= -i\sum_{ij}  {\bf R}_i [h_i,h_j].
\end{align}

By decomposing the energy current into a kinetic energy part and a potential energy part we get
\begin{align}
{\bf J}_{E}^{K.E.}
&=-i\sum_{i,\sigma}\sum_{{\bm \delta}{\bm \delta}^{\prime}} (t_{{\bm \delta}} t_{{\bm \delta}^{\prime}}/4)  \nonumber\\&\times
({\bf R}_{i+{\bm \delta}+{\bm \delta}^{\prime}}- {\bf R}_{i+{\bm \delta}}+ {\bf R}_{i+{\bm \delta}^{\prime}}- {\bf R}_{i})\nonumber\\&\times 
(c^{\dagger}_{i+{\bm \delta}+{\bm \delta}^{\prime} \sigma} c_{i\sigma}-c^{\dagger}_{i \sigma} c_{i+{\bm \delta}+{\bm \delta}^{\prime}\sigma}) .  
\end{align}

For the potential energy part, we obtain
\begin{align}
{\bf J}_{E}^{P.E.}&= -i\sum_{i,\sigma}\sum_{{\bm \delta}}  ( t_{{\bm \delta}} U/2) ({\bf R}_{i
+{\bm \delta}}-{\bf R}_i) \nonumber\\&\times(
c^{\dagger}_{i+{\bm \delta} \sigma}c_{i\sigma}-c^{\dagger}_{i \sigma}c_{i+{\bm \delta}\sigma})n_{i+{\bm \delta}\bar{\sigma}}.
\end{align}

\subsection{Correlation function} Having the explicit form of the current operators, $L^{12}(i\nu_n)$ coefficient can be written as
\begin{align}
L^{12}(i\nu_n)&=\frac{1}{\beta V}\int_0^{\beta}e^{i\nu_n\tau}\langle T_{\tau} {\bf J}(\tau)\cdot {\bf J}_Q(0) \rangle \nonumber\\
&=\frac{1}{\beta V}\int_0^{\beta}e^{i\nu_n\tau}\langle T_{\tau} {\bf J}(\tau)\cdot {\bf J}_{Q}^{K.E.}(0) \rangle\nonumber \\
&+\frac{1}{\beta V}\int_0^{\beta}e^{i\nu_n\tau}\langle T_{\tau} {\bf J}(\tau)\cdot {\bf J}_{Q}^{ P.E.}(0) \rangle.\label{eqJJ}
\end{align}

\subsubsection{Kinetic contribution} Since we neglect vertex corrections, we use Wick's theorem  for the kinetic energy contribution in $L_{12}$ coefficient, obtaining
\begin{align}
\frac{1}{\beta V}&\int_0^{\beta}e^{i\nu_n\tau}\langle T_{\tau} {\bf J}(\tau)\cdot {\bf J}_{Q}^{ K.E.}(0) \rangle=\nonumber\\
&\frac{1}{\beta V}\int_0^{\beta}e^{i\nu_n\tau}\sum_{ij\sigma}\sum_{{\bm \delta}{\bm \delta}^{\prime}{\bm \delta}^{\prime\prime}}(t_{{\bm \delta}}t_{{\bm \delta}^{\prime}}t_{{\bm \delta}^{\prime\prime}}/8)\nonumber\\
&({\bf R}_{i+{\bm \delta}}-{\bf R}_i)\cdot({\bf R}_{j+{\bm \delta}^{\prime}+{\bm \delta}^{\prime\prime}}- {\bf R}_{j+{\bm \delta}^{\prime}}+ {\bf R}_{j+{\bm \delta}^{\prime\prime}}- {\bf R}_{j})\nonumber\\
\bigg(&G_{j,i+{\bm \delta},\sigma}(-\tau)G_{i,j+{\bm \delta}^{\prime}+{\bm \delta}^{\prime\prime},\sigma}(\tau)\nonumber\\-&G_{j+{\bm \delta}^{\prime}+{\bm \delta}^{\prime\prime},i+{\bm \delta},\sigma}(-\tau)G_{i,j,\sigma}(\tau)\nonumber\\
-&G_{j,i,\sigma}(-\tau)G_{i+{\bm \delta},j+{\bm \delta}^{\prime}+{\bm \delta}^{\prime\prime},\sigma}(\tau)\nonumber\\
+&G_{j+{\bm \delta}^{\prime}+{\bm \delta}^{\prime\prime},i,\sigma}(-\tau)G_{i+{\bm \delta},j,\sigma}(\tau)\bigg),
\end{align}
where we neglect time independent terms. Transforming back to the Fourier space representation one finds
\begin{align}
\frac{1}{\beta V}&\int_0^{\beta}e^{i\nu_n\tau}\langle T_{\tau} {\bf J}^{\alpha}(\tau)\cdot {\bf J}^{\alpha, K.E.}_{Q}(0) =\nonumber\\
&-\frac{1}{\beta V}\sum_{{\bf k}\sigma}\sum_{\omega_m} \bigg({\bf v}^{\alpha}_{\bf k}G_{{\bf k}\sigma}(\omega_m)[\epsilon_{\bf k}-\mu]{\bf v}^{\alpha}_{\bf k}G_{{\bf k}\sigma}(\omega_m+\nu_n)\bigg),\label{eqJJKE}
\end{align}
where we used ${\bf v}_{-\bf k} = -{\bf v}_{\bf k}$ with ${\bf v}_{\bf k} =i\sum_{{\bm \delta}}{\bm \delta}t_{{\bm \delta}} \exp(i{\bf k}\cdot {\bm \delta})$.

\subsubsection{Potential contribution}  The potential energy contribution is
\begin{align}
\langle T_{\tau} &{\bf J}(\tau)\cdot {\bf J}_{Q}^{ P.E.}(0) \rangle=-\sum_{ij}\sum_{{\bm \delta}{\bm \delta}^{\prime}}\sum_{\sigma\sigma^{\prime}}(t_{{\bm \delta}} t_{{\bm \delta}^{\prime}}U/4) \nonumber\\ 
&\times [( {\bf R}_{i+{\bm \delta}}-{\bf R}_i)\cdot ( {\bf R}_{j+{\bm \delta}^{\prime}}-{\bf R}_j)]\nonumber\\
\bigg( &\langle T_{\tau} c^{\dagger}_{i+{\bm \delta} \sigma}(\tau) c_{i \sigma}(\tau)  
c^{\dagger}_{j+{\bm \delta}^{\prime} \sigma^{\prime}}(0)c_{j\sigma^{\prime}} (0)n_{j+{\bm \delta}^{\prime}\bar{\sigma}^{\prime}} (0)\rangle  \nonumber \\
-&\langle T_{\tau} c^{\dagger}_{i+{\bm \delta} \sigma}(\tau) c_{i \sigma}(\tau)  
c^{\dagger}_{j \sigma^{\prime}}(0)c_{j+{\bm \delta}^{\prime}\sigma^{\prime}} (0)n_{j+{\bm \delta}^{\prime}\bar{\sigma}^{\prime}} (0)\rangle  \nonumber \\
-&\langle T_{\tau} c^{\dagger}_{i\sigma}(\tau) c_{i+{\bm \delta}  \sigma}(\tau)  
c^{\dagger}_{j+{\bm \delta}^{\prime} \sigma^{\prime}}(0)c_{j\sigma^{\prime}} (0)n_{j+{\bm \delta}^{\prime}\bar{\sigma}^{\prime}} (0)\rangle  \nonumber \\
+&\langle T_{\tau} c^{\dagger}_{i \sigma}(\tau) c_{i+{\bm \delta} \sigma}(\tau)  
c^{\dagger}_{j \sigma^{\prime}}(0)c_{j+{\bm \delta}^{\prime}\sigma^{\prime}} (0)n_{j+{\bm \delta}^{\prime}\bar{\sigma}^{\prime}} (0)\rangle
\bigg).\label{eq:bk}
\end{align}

There are different ways to contract the expectation values in the Eq.~\ref{eq:bk} as a two points and a four points correlation function. For example the first braket on the right hand side of the Eq.~\ref{eq:bk} can be contracted as follows if we keep $c^{\dagger}_{i+{\bm \delta} \sigma}(\tau)$ in the two-point correlation functions:
\begin{align}
&\langle T_{\tau} c^{\dagger}_{i+{\bm \delta} \sigma}(\tau) c_{i \sigma}(\tau)\rangle \langle T_{\tau} 
c^{\dagger}_{j+{\bm \delta}^{\prime} \sigma^{\prime}}(0)c_{j\sigma^{\prime}} (0)n_{j+{\bm \delta}^{\prime}\bar{\sigma}^{\prime}} (0)\rangle\nonumber\\
+&\langle T_{\tau} c^{\dagger}_{i+{\bm \delta} \sigma}(\tau) c_{j\sigma^{\prime}} (0)\rangle \langle 
T_{\tau}c_{i \sigma}(\tau)  c^{\dagger}_{j+{\bm \delta}^{\prime} \sigma^{\prime}}(0)n_{j+{\bm \delta}^{\prime}\bar{\sigma}^{\prime}} (0)\rangle \nonumber\\
+&\langle T_{\tau} c^{\dagger}_{i+{\bm \delta} \sigma}(\tau) c_{j+{\bm \delta}^{\prime}\bar{\sigma}^{\prime}} (0)\rangle \langle 
T_{\tau}c_{i \sigma}(\tau)  c^{\dagger}_{j+{\bm \delta}^{\prime} \sigma^{\prime}}(0)
c_{j \sigma^{\prime}}(0) c^{\dagger}_{j+{\bm \delta}^{\prime} \bar{\sigma}^{\prime}}(0)
\rangle\label{Eq1B1}
\end{align}
 or as follows if we keep $c^{\dagger}_{j+{\bm \delta}^{\prime} \sigma^{\prime}}(0)$ in the two-point correlation functions:
\begin{align}
&\langle T_{\tau} c_{i \sigma}(\tau)  
c^{\dagger}_{j+{\bm \delta}^{\prime} \sigma^{\prime}}(0)\rangle \langle T_{\tau} c^{\dagger}_{i+{\bm \delta} \sigma}(\tau) c_{j\sigma^{\prime}} (0)n_{j+{\bm \delta}^{\prime}\bar{\sigma}^{\prime}} (0)\rangle
\nonumber\\
+&\langle T_{\tau}c^{\dagger}_{j+{\bm \delta}^{\prime} \sigma^{\prime}}(0)c_{j\sigma^{\prime}} (0)\rangle \langle T_{\tau} c^{\dagger}_{i+{\bm \delta} \sigma}(\tau) c_{i \sigma}(\tau)  n_{j+{\bm \delta}^{\prime}\bar{\sigma}^{\prime}} (0)\rangle \label{Eq1B2}
\end{align}
Since spin is conserved by the Hamiltonian, note that $\langle T_{\tau}c^{\dagger}_{j+{\bm \delta}^{\prime} \sigma^{\prime}}(0)c_{j+{\bm \delta}^{\prime}\bar{\sigma}^{\prime}} (0)\rangle=0$ due to opposite spins. 
We write the first bracket the Eq.~\ref{eq:bk} as summation of Eqs.~\ref{Eq1B1} and \ref{Eq1B2} divided by two. 

 The second braket on the right hand side of the Eq.~\ref{eq:bk} can be contracted as follows:
\begin{widetext}
\begin{align}
-\frac{1}{2}\bigg(&\langle T_{\tau} c^{\dagger}_{i+{\bm \delta} \sigma}(\tau) c_{i \sigma}(\tau)\rangle \langle T_{\tau} 
c^{\dagger}_{j \sigma^{\prime}}(0)c_{j+{\bm \delta}^{\prime}\sigma^{\prime}} (0)n_{j+{\bm \delta}^{\prime}\bar{\sigma}^{\prime}} (0)\rangle\nonumber\\
+&\langle T_{\tau} c^{\dagger}_{i+{\bm \delta} \sigma}(\tau) c_{j+{\bm \delta}^{\prime}\sigma^{\prime}} (0)\rangle \langle 
T_{\tau}c_{i \sigma}(\tau)  c^{\dagger}_{j \sigma^{\prime}}(0)n_{j+{\bm \delta}^{\prime}\bar{\sigma}^{\prime}} (0)\rangle \nonumber\\
+&\langle T_{\tau} c^{\dagger}_{i+{\bm \delta} \sigma}(\tau) c_{j+{\bm \delta}^{\prime}\bar{\sigma}^{\prime}} (0)\rangle \langle 
T_{\tau}
c_{i \sigma}(\tau)  c^{\dagger}_{j \sigma^{\prime}}(0)
c_{j+{\bm \delta}^{\prime} \sigma^{\prime}}(0)
c^{\dagger}_{j+{\bm \delta}^{\prime} \bar{\sigma}^{\prime}}(0)
\rangle\nonumber\\
+&\langle T_{\tau} c_{i \sigma}(\tau)  
c^{\dagger}_{j \sigma^{\prime}}(0)\rangle \langle T_{\tau} c^{\dagger}_{i+{\bm \delta} \sigma}(\tau) c_{j+{\bm \delta}^{\prime}\sigma^{\prime}} (0)n_{j+{\bm \delta}^{\prime}\bar{\sigma}^{\prime}} (0)\rangle
\nonumber\\
+&\langle T_{\tau} c^{\dagger}_{j \sigma^{\prime}}(0)c_{j+{\bm \delta}^{\prime}\sigma^{\prime}} (0)\rangle \langle T_{\tau}c^{\dagger}_{i+{\bm \delta} \sigma}(\tau) c_{i \sigma}(\tau)  n_{j+{\bm \delta}^{\prime}\bar{\sigma}^{\prime}} (0)\rangle
\bigg)\label{Eq2B}
\end{align}
\end{widetext}

 The first terms of Eqs.~\ref{Eq1B1} and  \ref{Eq2B} are time independent, so we neglect them. The third term in the Eqs.~\ref{Eq1B1}  can be contracted more to obtain
 \begin{align}
\delta_{\sigma,\bar{\sigma}^{\prime}} &\langle T_{\tau} c^{\dagger}_{i+{\bm \delta} \sigma}(\tau) c_{j+{\bm \delta}^{\prime}\sigma} (0)\rangle \langle 
T_{\tau}
c_{i \sigma}(\tau)  c^{\dagger}_{j+{\bm \delta}^{\prime} \sigma}(0)\rangle \nonumber \\
& \langle T_{\tau}
c^{\dagger}_{j+{\bm \delta}^{\prime} \bar{\sigma}}(0)
c_{j \bar{\sigma}}(0)\rangle.
\end{align}
This term cancels out with the third term in the Eq.~\ref{Eq2B} that has $\langle T_{\tau}
c^{\dagger}_{j \bar{\sigma}}(0)
c_{j+{\bm \delta}^{\prime} \bar{\sigma}}(0)\rangle$ instead of $\langle T_{\tau}
c^{\dagger}_{j+{\bm \delta}^{\prime} \bar{\sigma}}(0)
c_{j \bar{\sigma}}(0)\rangle$. These two expectation values are equal for a system in equilibrium. 
The last term in the Eqs.~\ref{Eq1B2}  is also cancelled out by the last term in the Eqs.~\ref{Eq2B} for a similar reason. 

Hence, the only surviving terms are the second, and fourth.  Keeping these terms, Eq.~\ref{eq:bk} becomes
\begin{align}
-\sum_{ij} &\sum_{{\bm \delta}{\bm \delta}^{\prime}}\sum_{\sigma}(t_{{\bm \delta}} t_{{\bm \delta}^{\prime}}U/8)  [( {\bf R}_{i+{\bm \delta}}-{\bf R}_i)\cdot ( {\bf R}_{j+{\bm \delta}^{\prime}}-{\bf R}_j)]\nonumber\\
\bigg( &\langle T_{\tau} c^{\dagger}_{i+{\bm \delta} \sigma}(\tau) c_{j\sigma} (0)\rangle \langle T_{\tau}n_{j+{\bm \delta}^{\prime}\bar{\sigma}} (0)
c_{i \sigma}(\tau)  
c^{\dagger}_{j+{\bm \delta}^{\prime} \sigma}(0)\rangle  \nonumber \\
+&\langle T_{\tau} c_{i \sigma}(\tau)  
c^{\dagger}_{j+{\bm \delta}^{\prime} \sigma}(0)\rangle \langle T_{\tau} n_{j+{\bm \delta}^{\prime}\bar{\sigma}} (0)c^{\dagger}_{i+{\bm \delta} \sigma}(\tau) c_{j\sigma} (0)\rangle
\nonumber\\
-&\langle T_{\tau} c^{\dagger}_{i+{\bm \delta} \sigma}(\tau) c_{j+{\bm \delta}^{\prime}\sigma} (0)\rangle \langle T_{\tau}
n_{j+{\bm \delta}^{\prime}\bar{\sigma}
} (0)c_{i \sigma}(\tau)  
c^{\dagger}_{j \sigma}(0)\rangle  \nonumber \\
-&\langle T_{\tau} c_{i \sigma}(\tau)  
c^{\dagger}_{j \sigma}(0)\rangle \langle T_{\tau} n_{j+{\bm \delta}^{\prime}\bar{\sigma}} (0)c^{\dagger}_{i+{\bm \delta} \sigma}(\tau) c_{j+{\bm \delta}^{\prime}\sigma} (0)\rangle
\nonumber\\
-&\langle T_{\tau} c^{\dagger}_{i\sigma}(\tau) c_{j\sigma} (0) \rangle \langle T_{\tau}
n_{j+{\bm \delta}^{\prime}\bar{\sigma}} (0)c_{i+{\bm \delta}  \sigma}(\tau)  
c^{\dagger}_{j+{\bm \delta}^{\prime} \sigma}(0)\rangle  \nonumber \\
-&\langle T_{\tau} c_{i+{\bm \delta}  \sigma}(\tau)  
c^{\dagger}_{j+{\bm \delta}^{\prime} \sigma}(0) \rangle \langle T_{\tau}
n_{j+{\bm \delta}^{\prime}\bar{\sigma}} (0)c^{\dagger}_{i\sigma}(\tau) c_{j\sigma} (0)
\rangle  \nonumber \\
+&\langle T_{\tau} c^{\dagger}_{i \sigma}(\tau) c_{j+{\bm \delta}^{\prime}\sigma} (0)\rangle \langle T_{\tau}n_{j+{\bm \delta}^{\prime}\bar{\sigma}} (0)
c_{i+{\bm \delta} \sigma}(\tau)  
c^{\dagger}_{j \sigma}(0)\rangle
\nonumber \\
+&\langle T_{\tau}c_{i+{\bm \delta} \sigma}(\tau)  
c^{\dagger}_{j \sigma}(0) \rangle \langle T_{\tau}n_{j+{\bm \delta}^{\prime}\bar{\sigma}} (0)c^{\dagger}_{i \sigma}(\tau) c_{j+{\bm \delta}^{\prime}\sigma} (0)
\rangle
\bigg)\label{eq01}
\end{align}

We use the following equation for the Hubbard model
 \begin{align}
-U\langle T_{\tau} &n_{i\bar{\sigma}}(\tau^{\prime})c_{i\sigma}(\tau^{\prime})c^{\dagger}_{j\sigma}(\tau)\rangle =\nonumber\\&
\int d\tau^{\prime\prime}\sum_l \Sigma_{il,\sigma}(\tau^{\prime}-\tau^{\prime\prime})G_{lj,\sigma}(\tau^{\prime\prime}-\tau)  \nonumber\\
 -U\langle T_{\tau} &n_{j\bar{\sigma}}(\tau^{\prime})c_{i\sigma}(\tau)c^{\dagger}_{j\sigma}(\tau^{\prime})\rangle=\nonumber\\&
 \int d\tau^{\prime\prime}\sum_l G_{il,\sigma}(\tau-\tau^{\prime\prime})\Sigma_{lj,\sigma}(\tau^{\prime\prime}-\tau^{\prime})  
 \end{align}
 to rewrite the first, fourth, fifth, and eighth term on the right hand side of the Eq.~\ref{eq01} as a multiplication of a Green's function and a self-energy. We conjecture that the remaining terms should decompose similarly.  To support this, consider another contraction on the four point correlation function of the second term. It gives 
 \begin{equation}
 (U\langle n_{j+{\bm \delta}^{\prime}\bar{\sigma}} \rangle)G_{j,i+{\bm \delta},\sigma}(-\tau)=\Re \Sigma_{{\bf k}\sigma}(\infty)G_{j,i+{\bm \delta},\sigma}(-\tau), 
 \end{equation}
 which is 
a multiplication of a Green's function and a self-energy.  In order to use same level of approximation for all terms in the right hand side of the Eq.~\ref{eq01}, we do not employ this latter extra contraction. Instead, we use the above equations for these terms as well.  We test numerically that this conjecture gives the right temperature dependence of the Seebeck effect for a system in Fermi liquid regime.

Therefore, we obtain
\begin{widetext}
\begin{align}
\sum_{ij,l}&\sum_{{\bm \delta}{\bm \delta}^{\prime}}\sum_{\sigma}\int d\tau^{\prime\prime}(t_{{\bm \delta}} t_{{\bm \delta}^{\prime}}/8)  [( {\bf R}_{i+{\bm \delta}}-{\bf R}_i)\cdot ( {\bf R}_{j+{\bm \delta}^{\prime}}-{\bf R}_j)]\nonumber\\
\bigg( &G_{i,l,\sigma}(\tau-\tau^{\prime\prime})\Sigma_{l,j+{\bm \delta}^{\prime},\sigma}(\tau^{\prime\prime}) G_{j,i+{\bm \delta},\sigma}(-\tau)
\nonumber+G_{i,j+{\bm \delta}^{\prime},\sigma}(\tau) \Sigma_{j,l,\sigma}(-\tau^{\prime\prime}) G_{l,i+{\bm \delta},\sigma}(\tau^{\prime\prime}-\tau) \nonumber\\
&-G_{j+{\bm \delta}^{\prime},i+{\bm \delta},\sigma}(-\tau)G_{i,l,\sigma}(\tau-\tau^{\prime\prime})\Sigma_{l,j,\sigma}(\tau^{\prime\prime}) \nonumber-\Sigma_{j+{\bm \delta}^{\prime},l,\sigma}(-\tau^{\prime\prime}) G_{l,i+{\bm \delta},\sigma}(\tau^{\prime\prime}-\tau)G_{i,j,\sigma}(\tau)
\nonumber\\
&- G_{i+{\bm \delta},l,\sigma}(\tau-\tau^{\prime\prime})\Sigma_{l,j+{\bm \delta}^{\prime},\sigma}(\tau^{\prime\prime}) G_{j,i,\sigma}(-\tau)-G_{i+{\bm \delta},j+{\bm \delta}^{\prime},\sigma}(\tau)\Sigma_{j,l,\sigma}(-\tau^{\prime\prime})G_{l,i,\sigma}(\tau^{\prime\prime}-\tau) \nonumber\\
&+G_{j+{\bm \delta}^{\prime},i,\sigma}(-\tau)G_{i+{\bm \delta},l,\sigma}(\tau-\tau^{\prime\prime}) \Sigma_{l,j,\sigma}(\tau^{\prime\prime})+\Sigma_{j+{\bm \delta}^{\prime},l,\sigma}(-\tau^{\prime\prime})G_{l,i,\sigma}(\tau^{\prime\prime}-\tau) G_{i+{\bm \delta},j,\sigma}(\tau)
\bigg).
\end{align}
\end{widetext}
Transforming back to the Fourier space representation one finds
the potential energy contribution as
\begin{align}
\frac{1}{\beta V}&\int_0^{\beta}e^{i\nu_n\tau}\langle T_{\tau} {\bf J}^{\alpha}(\tau)\cdot {\bf J}^{\alpha, P.E.}_{Q}(0) =-\frac{1}{2\beta V}\sum_{{\bf k}\sigma}\sum_{\omega_m}\nonumber\\&
\bigg( {\bf v}^{\alpha}_{{\bf k}}G_{{\bf k}\sigma}(\omega_m+\nu_n)[\Sigma_{{\bf k}\sigma}(\omega_m+\nu_n)/2\nonumber\\&+\Sigma_{{\bf k}\sigma}(\omega_m)/2]{\bf v}^{\alpha}_{{\bf k}} G_{{\bf k}\sigma}(\omega_m) + (\nu_n \leftrightarrow -\nu_n)\bigg),\label{eqJJPE}
\end{align}

The summation of Eq.~\ref{eqJJKE} and Eq.~\ref{eqJJPE} gives the final results for the $L^{12}(i\nu_n)$ as
\begin{align}
L^{12}_{\alpha\alpha}&(i\nu_n)=-\frac{1}{2\beta V}\sum_{{\bf k}\sigma}\sum_{\omega_m}
\bigg( {\bf v}^{\alpha}_{{\bf k}}G_{{\bf k}\sigma}(\omega_m+\nu_n)\nonumber\\&\times [\epsilon_{\bf k}-\mu+\Sigma_{{\bf k}\sigma}(\omega_m+\nu_n)/2+\Sigma_{{\bf k}\sigma}(\omega_m)/2]\nonumber \\& {\bf v}^{\alpha}_{{\bf k}} G_{{\bf k}\sigma}(\omega_m) + (\nu_n \leftrightarrow -\nu_n).
\bigg),\label{eqL12}
\end{align}
%
Note that the term resulting from $\nu_n \leftrightarrow -\nu_n$ is equal to the term written explicitly. However, in numerical calculations with a finite frequency cutoff it is better to use the above form. 


\subsection{Analytic continuation: result} Analytic continuation can be done by employing the spectral representation of the Green's function and the self-energy
\begin{align}
G_{{\bf k}\sigma}(i\omega_m) &= \int d\omega \frac{A_{{\bf k}\sigma}(\omega)}{(i\omega_m-\omega)},\\
\Sigma_{{\bf k}\sigma}(i\omega_m)- \Re \Sigma(\infty)&= \int d\omega \frac{B_{{\bf k}\sigma}(\omega)}{(i\omega_m-\omega)},
\end{align}
where $A_{{\bf k}\sigma}(\omega)=(-1/\pi)\Im G_{{\bf k}\sigma}(\omega)$ and $B_{{\bf k}\sigma}(\omega)=(-1/\pi)\Im \Sigma_{{\bf k}\sigma}(\omega)$. 

%

We obtain $ \Im L^{12}(\nu)$
\begin{widetext}
\begin{align}
\Im L^{12}(\nu)=&-\frac{\pi}{ V}\sum_{{\bf k}\sigma}\int d\omega
\bigg( {\bf v}^{\alpha}_{{\bf k}}A_{{\bf k}\sigma}(\omega+\nu)[\epsilon_{\bf k}-\mu+\Re \Sigma_{{\bf k}\sigma}(\omega+\nu)/2+\Re \Sigma_{{\bf k}\sigma}(\omega)/2]{\bf v}^{\alpha}_{{\bf k}} A_{{\bf k}\sigma}(\omega)
\bigg)(n_F(\omega+\nu)-n_F(\omega))
\nonumber\\
&-\frac{\pi}{2V}\sum_{{\bf k}\sigma}\int d\omega \bigg( {\bf v}^{\alpha}_{{\bf k}}\Re G_{{\bf k}\sigma}(\omega+\nu)B_{{\bf k}\sigma}(\omega+\nu) {\bf v}^{\alpha}_{{\bf k}} A_{{\bf k}\sigma}(\omega)\bigg)(n_F(\omega+\nu)-n_F(\omega))\nonumber\\
&-\frac{\pi}{2V}\sum_{{\bf k}\sigma}\int d\omega \bigg( {\bf v}^{\alpha}_{{\bf k}}A_{{\bf k}\sigma}(\omega+\nu)B_{{\bf k}\sigma}(\omega) {\bf v}^{\alpha}_{{\bf k}} \Re G_{{\bf k}\sigma}(\omega)\bigg)(n_F(\omega+\nu)-n_F(\omega)),\label{L12Im}
\end{align}
\end{widetext}
where we used the Kramers-Kroning relation for the Green's function and the self-energy, i.e.,
\begin{align}
\Re G_{{\bf k}\sigma}(\omega)&={\mathcal P}  \int d\omega^{\prime} \frac{A_{{\bf k}\sigma}(\omega^{\prime})}{\omega-\omega^{\prime}},\\
\Re \Sigma_{{\bf k}\sigma}(\omega)-\Re\Sigma(\infty)&={\mathcal P}  \int d\omega^{\prime} \frac{B_{{\bf k}\sigma}(\omega^{\prime})}{\omega-\omega^{\prime}}
\end{align}
Note that $\Im L^{12}(\nu)$ is an odd function of $\nu$.  Finally, one can obtaine $ \mathcal{A}^{12}(\nu) \equiv \Im L^{12}(\nu)/\nu$ by dividing the Eq.~\ref{L12Im} by $\nu$. 

\subsection{Derivation of the analytic continuation formula}
Equation~\ref{L12Im} can be derived as follows. After replacing spectral representations of the Green's function and self energy in Eq.~\ref{eqL12}, we encounter the following integrals
\begin{align}
\int d\omega_1\; d\omega_2 \; d\omega_3 &A_{{\bf k}\sigma}(\omega_1)B_{{\bf k}\sigma}(\omega_2)A_{{\bf k}\sigma}(\omega_3) \nonumber \\ &\times S_i(\nu,\omega_1,\omega_2,\omega_3), \;\; (i=1,2)
\end{align}
where on the imaginary axis
\begin{widetext}
\begin{align}
S_1(i\nu_n,&\omega_1,\omega_2,\omega_3)=\frac{1}{\beta}\sum_{\omega_m} \frac{1}{i(\omega_m+\nu_n)-\omega_1}\nonumber \frac{1}{i(\omega_m+\nu_n)-\omega_2} \cdot \frac{1}{i\omega_m-\omega_3},\\
S_2(i\nu_n,&\omega_1,\omega_2,\omega_3)=\frac{1}{\beta}\sum_{\omega_m} \frac{1}{i(\omega_m+\nu_n)-\omega_1}\nonumber \frac{1}{i\omega_m-\omega_2} \cdot \frac{1}{i\omega_m-\omega_3}
\end{align}

The summation over fermionic Matsubara frequency can be done as follows: If we write $S(i\nu_n) = (\frac{1}{\beta})\sum_{\omega_m} F(\omega_m)$, then $S(i\nu_n) = \sum_{i} r_i n_F(z_i)$, where $z_i$ is a simple pole of $F(z)$, $r_i$ is the corresponding residue and $n_F$ is Fermi-Dirac distribution with, as usual, $n_F(\omega \pm i\nu_n) = n_F(\omega)$. Therefore, we obtain
\begin{align}
S_1(i\nu_n, &\omega_1,\omega_2,\omega_3) = (\frac{1}{\omega_1-\omega_2})(\frac{1}{\omega_1-i\nu_n-\omega_3})\;n_F(\omega_1)\nonumber\\&+ (\frac{1}{\omega_2-\omega_1})(\frac{1}{\omega_2-i\nu_n-\omega_3})\;n_F(\omega_2)
+ (\frac{1}{\omega_3+i\nu_n-\omega_1})(\frac{1}{\omega_3+i\nu_n-\omega_2})\;n_F(\omega_3),\\
S_2(i\nu_n, &\omega_1,\omega_2,\omega_3) = (\frac{1}{\omega_1-i\nu_n-\omega_2})(\frac{1}{\omega_1-i\nu_n-\omega_3})\;n_F(\omega_1)\nonumber\\&+ (\frac{1}{\omega_2+i\nu_n-\omega_1})(\frac{1}{\omega_2-\omega_3})\;n_F(\omega_2)
+ (\frac{1}{\omega_3+i\nu_n-\omega_1})(\frac{1}{\omega_3-\omega_2})\;n_F(\omega_3).
\label{eq:MS}
\end{align}

Note that in the case  $\omega_1=\omega_2$ for $S_1$, the $\omega_1$ pole is a second-order pole. Nevertheless, the summation of the first two terms on the right hand side give the correct result for a second order pole. A similar argument is valid for $S_2$.
The next step is to convert to a retarded function ($i\nu_n \rightarrow \nu+i0^+$).
Employing the identity, 
\begin{equation}
\frac{1}{\omega-\omega_0\pm i0^+} = {\mathcal P} \frac{1}{\omega-\omega_0}\mp i\pi\delta(\omega-\omega_0),
\end{equation}
the imaginary parts of $S_i(\nu)$ are 
\begin{align}
\Im S_1&(\nu, \omega_1,\omega_2,\omega_3) = 
(\frac{\pi n_F(\omega_1)}{\omega_1-\omega_2})\delta(\omega_1-\nu-\omega_3)+ (\frac{\pi n_F(\omega_2)}{\omega_2-\omega_1})\delta(\omega_2-\nu-\omega_3)\nonumber\\
- &{\mathcal P} (\frac{\pi n_F(\omega_3)}{\omega_3+\nu-\omega_1})\delta(\omega_3+\nu-\omega_2)
- {\mathcal P} (\frac{\pi n_F(\omega_3)}{\omega_3+\nu-\omega_2})\delta(\omega_3+\nu-\omega_1),\\
\Im S_2&(\nu, \omega_1,\omega_2,\omega_3) = 
{\mathcal P} (\frac{\pi n_F(\omega_1)}{\omega_1-\nu-\omega_2})\delta(\omega_1-\nu-\omega_3)+ {\mathcal P} (\frac{\pi n_F(\omega_1)}{\omega_1-\nu-\omega_3})\delta(\omega_1-\nu-\omega_2)\nonumber\\- &(\frac{\pi n_F(\omega_2)}{\omega_2-\omega_3}\delta(\omega_2+\nu-\omega_1)- (\frac{\pi n_F(\omega_3)}{\omega_3-\omega_2})\delta(\omega_3+\nu-\omega_1)
\label{eq:MSReal}
\end{align}

\end{widetext}
%


%

\end{document}